\documentclass[aps,preprint,tightenlines,showpacs,nofootinbib]{revtex4}

\usepackage{graphicx}  
\usepackage{bm}  
\usepackage{amsmath}
\newcommand{\beq}{\begin{equation}}
\newcommand{\eeq}{\end{equation}}
\newcommand{\bea}{\begin{eqnarray}}
\newcommand{\eea}{\end{eqnarray}}

\begin{document}
\title{Model-independent extraction of two-photon effects\\
in elastic electron-proton scattering\\
\vspace{-3cm}\hfill{\tiny HISKP-TH-07/14, FZJ-TH-2007-16}\vspace{3cm} 
}
\author{M.A. Belushkin}\email{belushki@itkp.uni-bonn.de}
\author{H.-W. Hammer}\email{hammer@itkp.uni-bonn.de}
\affiliation{Helmholtz-Institut f\"ur Strahlen- und Kernphysik (Theorie),
Universit\"at Bonn, Nu\ss allee 14-16,\\ D-53115 Bonn, Germany}

\author{Ulf-G. Mei\ss ner}\email{meissner@itkp.uni-bonn.de}
\affiliation{Helmholtz-Institut f\"ur Strahlen- und Kernphysik (Theorie),
Universit\"at Bonn, Nu\ss allee 14-16, D-53115 Bonn, Germany\\
and\\
Institut f\"ur Kernphysik (Theorie), Forschungszentrum
J\"ulich, D-52425 J\"ulich, Germany\\}

\date{\today}
\begin{abstract}
We address the discrepancy between the Rosenbluth and polarization
transfer data for the electromagnetic form factors of the nucleon.
Assuming that the effect of two-photon corrections on the 
polarization transfer data is negligible, we obtain a
model-independent estimate of the two-photon correction $\Delta^{2\gamma}$.
We analyze the polarization transfer data and the cross 
section data separately using dispersion relations. A central value as well as an error  
estimate for $\Delta^{2\gamma}$ is then obtained from a comparison of the two analyses.
The resulting values for $\Delta^{2\gamma}$ are in good agreement with
direct calculations available in the literature.
\end{abstract}
\maketitle
{\em Introduction}~---
The electromagnetic form factors of the nucleon encode information about the nucleon structure as probed by the
electromagnetic interaction \cite{Gao:2003ag,Hyde-Wright:2004gh,Perdrisat:2006hj}.
At low momentum transfers, the electric form factors ($G_{E}$) and the magnetic form factors
($G_{M}$) can be thought of as a Fourier transform of the charge and magnetisation distributions within the nucleon,
respectively. Two experimental techniques used to extract the nucleon form factors from elastic electron-nucleon
scattering exist: the Rosenbluth and the polarization transfer methods. The Rosenbluth technique is used to extract
the form factors from elastic cross section data for unpolarised electron-nucleon scattering. The Rosenbluth formula for
the differential cross section \cite{Rosenbluth:1950} is given by the product of the Mott cross section, which
corresponds to scattering on a point-like particle, and a linear combination of the squared nucleon form factors:
\begin{equation}\label{eq:xs_ros}
\frac{d\sigma}{d\Omega} = \left( \frac{d\sigma}{d\Omega}\right)_{\rm Mott} \frac{\tau}{\epsilon (1+\tau)}
\left[G_{M}^{2}(Q^{2}) + \frac{\epsilon}{\tau} G_{E}^{2}(Q^{2})\right]\, ,
\end{equation}
where $\epsilon = [1+2(1+\tau)\tan^{2} (\Theta/2)]^{-1}$ is the virtual photon polarization,
$\Theta$ is the electron scattering angle in the
laboratory frame, and $\tau = Q^2/4M_{N}^{2}$, with $Q^{2}$ the (negative of the) invariant four-momentum transfer squared and
$M_{N}$ the nucleon mass. Because the form factors in Eq.~(\ref{eq:xs_ros}) are functions of $Q^2$
only, it is possible to extract $G_{M}$ and the ratio $G_{E}/G_{M}$ from studying the $\epsilon$ dependence
of the cross sections at a fixed value of $Q^{2}$.
Because of the $1/\tau$ prefactor of $G_{E}$, however, the cross section becomes increasingly dominated by $G_{M}$ at
larger $Q^{2}$ values, and the extraction of the electric form factor becomes difficult.
\par
An alternative method for the measurement of the nucleon form factors has been actively developed since
1968 - the polarization transfer method (PT) \cite{Akhiezer:1968ek,Akhiezer:1974em,Arnold:1980zj,Dombey:1969wk}.
It gives high sensitivity in the extraction of the ratio $G_{E}/G_{M}$, and features significantly
smaller systematic uncertainties. In 1999 and 2001, the results of high precision measurements of this ratio 
at Jefferson Lab were presented~\cite{Jones:1999rz,Gayou:2001qd}. These results were found to be in striking
disagreement with the world data from Rosenbluth separation. This discrepancy has been a subject of intense investigation
in recent years, raising questions about the validity of these techniques.
\par
When the discrepancy between the Rosenbluth and the polarization transfer measurement results was first observed,
it was noted that the values of $G_{E}^{p}$ extracted using the Rosenbluth technique are not consistent with each other.
It was often assumed that the difference between the results of the two techniques can therefore be explained by
systematic uncertainties in the Rosenbluth extractions. A global reanalysis of the Rosenbluth data~\cite{Arrington:2003df}
confirmed that only the data from individual experiments are consistent. The results of the reanalysis still failed to
bring agreement with the polarization transfer data. 
\par
Another possible reason for the discrepancy was found in the radiative corrections. Mo and Tsai \cite{Mo:1968cg} 
and earlier Meister and Yennie \cite{Meister:1962} calculated
higher order corrections to the Rosenbluth formula including all standard QED radiative corrections up
to ${\cal O}(\alpha^2)$ with $\alpha\approx 1/137$ the fine structure constant.  
These corrections are the ones applied to most experimental data. The
calculations, however, involved a number of approximations. In
particular, the effects of the structure of the nucleon were neglected in the calculation of the two-photon box and
crossed-box diagrams. This approximation has been a subject of intense investigation in recent years, and several
model calculations of the full contribution including the nucleon structure
have been performed. In Ref.~\cite{Blunden:2003sp}, the authors calculated
the two-photon box and crossed-box graphs directly using Feynman diagram techniques. The calculation considered an elastic
proton intermediate state, with the form factors parameterised by a simple monopole. 
Subsequently, this calculation was improved by using a better parameterisation of the proton 
form factors with a sum of monopoles~\cite{Blunden:2005ew} and by including the $\Delta$
intermediate state~\cite{Kondratyuk:2005kk}. Another approach to calculating the two-photon exchange contributions was
taken in Refs.~\cite{Chen:2004tw,Afanasev:2005mp} within the framework of generalized parton
distributions. Both approaches lead to sizeable contributions 
that reduce the discrepancy between the Rosenbluth and the polarization transfer data.
However, they fail to bring complete agreement between the two techniques when applied to 
the form factor ratios.
\par
Another important source of corrections are static Coulomb effects. They arise from to the distortion of
the electron wave in the Coulomb field of the proton. In terms of Feynman diagrams, this corresponds to the exchange
of one hard and one (or several) soft photons between the electron and the proton. 
We have applied the Coulomb corrections obtained in second order Born approximation directly to the experimental data for the differential cross section. At this order, the dominant contribution from the exchange of one hard and one soft photons
is considered. This is similar to what was done in Ref.~\cite{Arrington:2004is}. 
The numerical values for the Coulomb corrections were taken from \cite{Sick:private}.
Note also that these Coulomb corrections are a well known subset of the two-photon corrections calculated in 
Refs.~\cite{Blunden:2003sp,Blunden:2005ew,Kondratyuk:2005kk}.

\medskip

{\em Cross section analysis}~---
In this paper, we have taken an alternative approach aimed at estimating the two-photon
corrections not present in the treatment of Mo and Tsai in a model-independent way. 
It has been shown that the polarization transfer data are not affected by two-photon corrections 
significantly~\cite{Blunden:2005ew}. Based on this fact, we have estimated the two-photon (and possible other) corrections 
not present in the treatment of Mo and Tsai~\cite{Mo:1968cg} as follows.
First, we have performed a global analysis of the Rosenbluth cross section data for the proton with the corrections of Mo and Tsai 
as well as Coulomb corrections \cite{Sick:private} 
applied and the neutron form factor data. 
The results of this new analysis were compared to the results of our previous analysis~\cite{Belushkin:2006qa} 
which included the polarization data but not the contradictory Rosenbluth data above $Q^2\simeq 1$ GeV$^2$.
For the reason discussed above,
the difference between the two analyses gives an estimate of the non-standard two-photon effects.

This comparison is easiest to carry out at the level of cross sections.
We reconstruct the differential cross section using our analysis of the polarization
transfer data for the form factors using
\begin{equation}\label{eq:xs_pol_ff}
\left( \frac{d\sigma}{d\Omega}\right)_{\mathrm A} = \left ( \frac{d\sigma}{d\Omega} \right)_{\rm Mott}
\frac{G_{E}^{2} + \tau G_{M}^{2}}{1+\tau} + 2\tau G_{M}^{2} \tan^{2} (\Theta/2)\,,
\end{equation}
where A = SC, pQCD indicates the treatment of constraints from perturbative QCD in the dispersion 
relations: 
SC uses superconvergence relations only, while pQCD includes an additional explicit pQCD term~\cite{Belushkin:2006qa}.
Both approaches give similar results for the form factors. We assume the reconstructed 
cross section Eq.~(\ref{eq:xs_pol_ff}) to be the Born cross section free of two-photon effects. Moreover,
we relate the cross section obtained from the new dispersion analysis of unpolarised elastic electron-proton cross sections
with the standard corrections of Mo and Tsai as well as the Coulomb corrections applied
to the cross section of Eq.~(\ref{eq:xs_pol_ff}) in the usual way \cite{Schwinger:1949ra},
\begin{equation}\label{eq:xs_to_xs_pol}
\left( \frac{d\sigma}{d\Omega}\right)_{\rm Ros,A} = 
\left( \frac{d\sigma}{d\Omega}\right)_{\rm A} (1+\delta^{2\gamma})\,.
\end{equation}
Here $\delta^{2\gamma}$ includes the corrections due to the exchange of two hard photons, as well as possible
higher-order corrections not present in the calculation of Mo and Tsai. 
\par
The direct calculation of
the two-photon corrections of Refs.~\cite{Blunden:2003sp,Blunden:2005ew,Kondratyuk:2005kk} includes the Coulomb
corrections as a subset of the two-photon box diagram. In order to compare our results to the results of Blunden {\it et al.}
\cite{Blunden:2005ew} below, we must therefore add the Coulomb corrections $\delta^{C}$  \cite{Sick:private}
to our extraction of the hard two-photon corrections $\delta^{2\gamma}$,
\begin{equation}\label{eq:def_Delta_by_delta}
\Delta^{2\gamma} = \delta^{2\gamma} + \delta^{C}\,.
\end{equation}
\par
We proceed to estimate the hard two-photon correction  $\delta^{2\gamma}$ using the form factors of Ref.~\cite{Belushkin:2006qa} 
to reconstruct the cross section Eq.~(\ref{eq:xs_pol_ff}) on the right hand side in Eq.~(\ref{eq:xs_to_xs_pol}) 
and our new analysis of the unpolarised cross section data as follows:
\begin{equation}\label{2gamma_xs}
\delta^{2\gamma} = \left( \frac{d\sigma}{d\Omega} \right) _{\rm Ros,A} \Bigg /
\left( \frac{d\sigma}{d\Omega} \right)_{\rm A} -1\,,
\end{equation}
where A = SC, pQCD indicates the treatment of the perturbative QCD constraints. 
We note that $\delta^{2\gamma}$ extracted this way might contain other higher-order corrections. However, we expect the 
hard two-photon corrections to dominate.

\medskip

{\em Results}~---
To estimate the sensitivity to the specific parametrization of the pQCD behavior, we carry out the analysis both in the
SC and pQCD approaches.
The previous fit to form factor data within the SC approach~\cite{Belushkin:2006qa} has 17 free parameters and
a total $\chi^{2}/$dof of 1.8. The previous fit to form factor data within the pQCD approach~\cite{Belushkin:2006qa} has
14 free parameters and a total $\chi^{2}/$dof of 2.0.
Our current fit to cross section data within the SC approach
has 14 free parameters and a total $\chi^{2}/$dof of 1.4. However, the biggest contribution to the 
$\chi^{2}/$dof in this fit comes from the time-like data. 
The $\chi^{2}/$dof for space-like data only (cross section data and the neutron form factor data) of this fit is 0.86.
The current fit to cross section data within the pQCD approach
has 20 free parameters and a total $\chi^{2}/$dof of 1.14.
The $\chi^{2}/$dof for space-like data only (cross section data and the neutron form factor data) is 0.91, slightly higher
than in the SC approach. Both fits give a similar description of the experimental data.
\par
\begin{figure}[ht]
\begin{center}
\includegraphics[width=14cm]{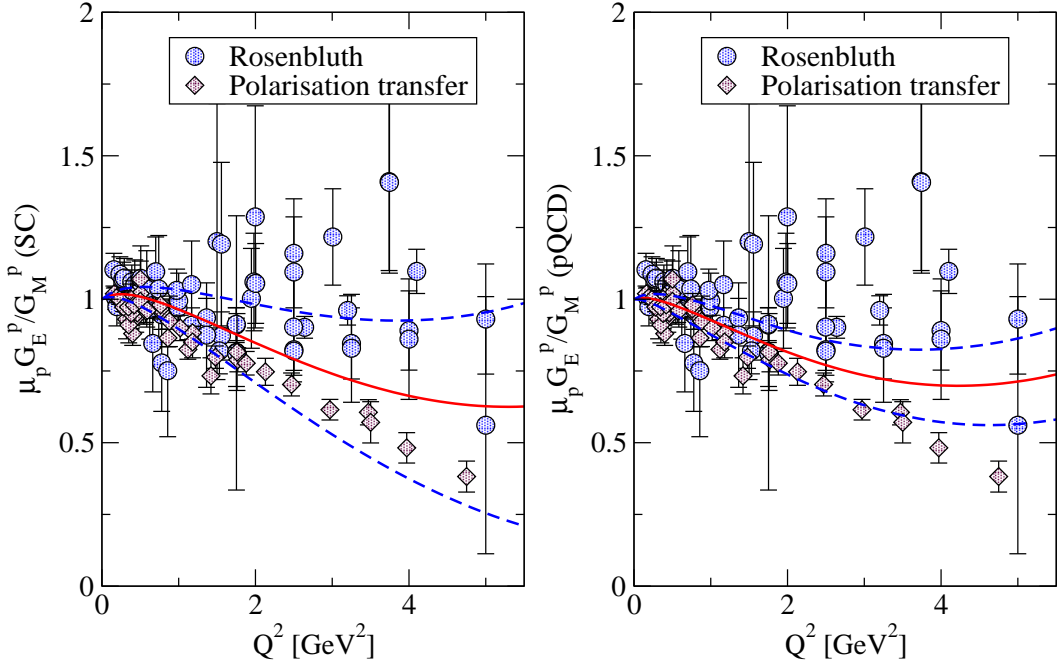}
\end{center}
\caption{Results obtained for the ratio $\mu_{p}G_{E}^{p}/G_{M}^{p}$ in the cross section analysis within the SC 
approach (left panel) and the pQCD approach (right panel) compared to the world data. The circles correspond to world data 
extracted using the Rosenbluth technique. The diamonds show the polarization data from Jefferson 
Lab~\cite{Jones:1999rz,Gayou:2001qd}. The form factor data shown on the plot did not participate in the fits. 
The solid lines show the best fits while the dashed lines give the theoretical $1\sigma$ uncertainty 
bands. 
\label{fig:res_sc_gegm}}
\end{figure}
In Fig.~\ref{fig:res_sc_gegm}, we show the ratio $\mu_{p}G_{E}^{p}/G_{M}^{p}$ extracted from the cross section
analysis within the SC approach (left panel) and the pQCD approach (right panel) compared to the world data.
We note that the form factor data shown in the plots did not participate in
the fits since we directly fitted to the cross sections. 
The solid lines show the best fits while the dashed lines give the theoretical $1\sigma$ uncertainty bands.
The large theoretical uncertainty demonstrates the insensitivity of the cross section data to $G_{E}^{p}/G_{M}^{p}$ 
within the experimental errors.
For the SC results, the polarization transfer data lie within the $1\sigma$ band, while some of the world 
Rosenbluth data do not. The $1\sigma$ uncertainty in the pQCD approach is slightly smaller than in the SC approach 
which is due to the particular configuration of the parameter space in the vicinity of the minimum. 
Both results agree very well within the uncertainty bands.
\par
\begin{figure}[ht]
\begin{center}
\includegraphics[width=15cm]{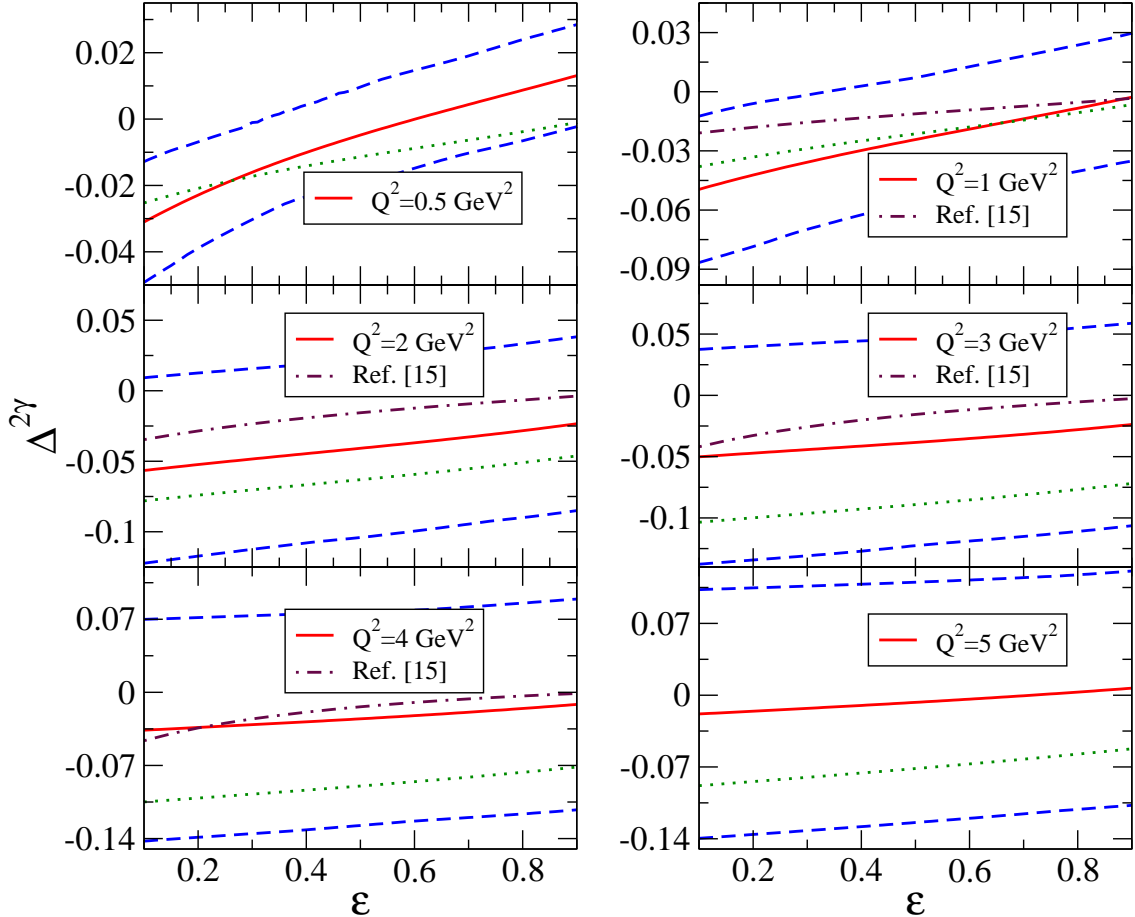}
\end{center}
\caption{The extracted two-photon correction $\Delta^{2\gamma}$ for $Q^2=$0.5, 1.0, 2.0, 3.0, 4.0,
and 5.0 GeV$^2$. Our correction $\Delta^{2\gamma}$ corresponds to the sum of the extracted hard two-photon
correction $\delta^{2\gamma}$ and the Coulomb correction $\delta^C$.
The solid lines show our results for $\Delta^{2\gamma}$ within the SC approach, while the dashed 
lines gives the theoretical $1\sigma$ uncertainty bands. The dotted lines give the central value for $\Delta^{2\gamma}$ 
within the pQCD approach. For comparison, the direct full calculation by Blunden {\it et al.}~\cite{Blunden:2005ew}
is shown by the dash-dotted lines.
\label{fig:res_2gamma}}
\end{figure}
The values of $\Delta^{2\gamma}=\delta^{2\gamma}+\delta^C$ 
extracted from the comparison of the cross section analysis and the 
form factor analysis~\cite{Belushkin:2006qa} are shown in Fig.~\ref{fig:res_2gamma} for $Q^2=$0.5, 1.0, 2.0, 3.0, 4.0,
and 5.0 GeV$^2$. The solid line corresponds to the extracted $\Delta^{2\gamma}$ in the SC approach, while
the dashed lines indicate the $1\sigma$ uncertainty bands. The theoretical uncertainty
arises from the error propagation in the ratio of Eq.~(\ref{2gamma_xs}).
For comparison we also show the direct calculation of Blunden {\it et al.}~\cite{Blunden:2005ew} by the dash-dotted
lines. The results of the SC approach are in excellent agreement with this calculation. 
The pQCD result is in very good agreement with the SC result up to $Q^{2} \approx 2$ GeV$^{2}$, but predicts a larger
$\Delta^{2\gamma}$ at higher $Q^{2}$ values. However, both approaches agree
well within the theoretical error bands.

\medskip

{\em Summary and conclusions}~---
We have performed a model-independent estimate of the additional two-photon corrections not present in the treatment of 
Mo and Tsai~\cite{Mo:1968cg}. We observed that
at least up to $Q^{2}\approx$ 4 GeV$^{2}$, the difference between the  form factors extracted using the 
Rosenbluth and the polarization transfer techniques can be explained by the two-photon
exchange corrections calculated in Ref.~\cite{Blunden:2005ew} well within the experimental error bars. 
The standard corrections of Mo and Tsai, the additional hard two-photon exchange corrections and the Coulomb corrections,
however, must be applied to the original cross section data.
Another source of corrections to consider are mathematical approximations and corrections due to the nucleon size effects
which give a further 1\% level contribution~\cite{Maximon:2000hm}. These corrections become increasingly important at
high $Q^{2}$ values.
\par
In order to apply the corrections to the form factors
themselves, a global reanalysis of the unpolarised elastic electron-proton differential cross section data should be performed.
Moreover, special care must be taken to study the influence of variations in $G_{E}^{p}$ on the overall analysis. 
In summary, we have shown
that meaningful results in much better agreement with the polarization transfer data can be obtained from the
current world data for unpolarised cross sections if the standard corrections of Mo and Tsai are applied consistently
together with the additional hard two-photon exchange corrections and the Coulomb corrections. 

\medskip

{\em Acknowledgements}~---
We thank W. Melnichouk for useful discussions and  for providing his results
for the two-photon corrections.
We thank John Arrington and Ingo Sick for valuable discussions of the Coulomb corrections.
We also thank Ingo Sick for providing the database for the elastic electron-proton cross sections
and the Coulomb corrections.  This work was supported in part by the EU I3HP
\lq\lq Study of Strongly Interacting Matter'' under contract No.
RII3-CT-2004-506078, by the BMBF under contract No. BN06BN411, 
by the  EU Contract No. MRTN-CT-2006-035482, \lq\lq FLAVIAnet'', and
by the DFG through funds provided
to the SFB/\-TR 16 \lq\lq Subnuclear Structure of Matter''.


\end{document}